\font\bigtitle=cmbx12 at 14pt
\font\title=cmbx12
\font\usual=cmr12
\rightline{{CERN-TH/97-140}}
\rightline{{hep-th/9706208}}

\usual

\bigskip
\bigskip
\bigskip
{\centerline{{\bigtitle{SL(2,Z) duality of Born-Infeld theory from
non-linear}}}
{\centerline{{\bigtitle{self-dual electrodynamics in 6 dimensions}}}
\bigskip
\bigskip
\bigskip
{\centerline{{David Berman}}\medskip

 {\centerline{ Dept. Mathematics, Durham University, South Road, Durham,
UK}} \smallskip {\centerline{ email: D.S.Berman@dur.ac.uk}}} \bigskip
\bigskip \bigskip 
\bigskip
\bigskip\bigskip\bigskip\bigskip
{\centerline{{\title{Abstract}}} \bigskip

We reformulate the Born-Infeld action, coupled to an axion and a dilaton
in a duality manifest way. This action is the generalization of the
Schwarz-Sen action for non-linear electrodynamics. We show that this
action may be obtained by dimensional reduction on a torus of a self-dual
theory in 6 dimensions. The dilaton-axion being identified with the
complex structure of the torus. Applications to M-theory and the
self-dual IIB three brane are investigated.

\vfill
{{CERN-TH/97-140}}

{{hep-th/9706208}}

June 1997
\eject

\baselineskip=18pt

{\bigtitle{Introduction}}
\bigskip

Recently, there has been much interest in the role of dualities in string
and field theories. Usually, these duality symmetries are not manifest in
the action but are seen as symmetries of the equations of motion together
with the Bianchi Identities. In the most elementary example, of
4-dimensional electromagnetism in the absence of sources, the equations of
motion and the Bianchi Identities swap roles, hence electric and magnetic
fields, are exchanged. 
Specifically, $E \rightarrow B$ and $ B \rightarrow -E$. Yet, the action
is not invariant under such a transformation. With the contempory
understanding of the importance of duality, it is desirable to incorporate
these hidden duality symmetries directly into the action. In [1], Schwarz
and Sen describe how to form duality invariant actions for, amongst 
others, the Maxwell action and the low energy effective action of
heterotic string theory compactified on a six torus. An essential
ingredient in their formulation was to give up manifest Lorentz
invariance. This was inspired by the work of various people [2] in writing
down the action for self-dual field theories. Later, it was shown that
manifest Lorentz invariance may be recovered by including in the action an
auxiliary field that allows some non trivial gauge symmetries. This is the
so called PST formalism [3].

One of the most major developments in recent years is the introduction of
D-branes in string theory [4]. It has been shown that the world volume
action of a D-brane is of Born-Infeld type [4,5]. In particular, the
bosonic D-3 brane will have a world volume action that is the
Born-Infeld action. It has been shown, [5,6] that this action has a
duality symmetry of a similar type to the Maxwell theory. Under world
volume duality transformation of the vector fields in the brane, the form
of the action is left invariant but the background axion-dilaton fileds
are inverted. This may be
extended to give a full SL(2,R) duality once shifts to the
axion field are taken into account. Of course we only expect the duality
group to be SL(2,Z) for the quantum theory.

 We shall describe how one may produce the 4-dimensional Born-Infeld
action associated with the 3-brane from self-dual non-linear 2-form
electrodynamics in 6-dimensions, via dimensional reduction on a torus.
This is the theory associated with the 5-brane in M-theory [7]. In
fact it is this theory, which has only been recently described, that will
prove vital to our construction. Moreover, we will in the process
reformulate the Born-Infeld theory in such away that the duality symmetry
is manifest- in a generalization of the Schwarz-Sen construction discussed
above.  Importantly, the duality in the four dimensional theory will be
shown to be a consequence of the geometry of the torus. That is we shall
be able to identify the coupling in the theory with the complex structure
of the torus. Duality then results from transformations of the complex
structure that leave the torus invariant ie $SL(2,Z)$. This is in the
spirit of [8] where duality in lower dimensional theories is seen as a
geometrical property of the compact space used in dimensional reduction. 
Of course, the context for this work is in M-theory. The Supersymmetric
version of these actions should be associated with the M-theory 5-brane
and the IIB self-dual 3-brane. In that context, the SL(2,Z) symmetry is an
S-duality in IIB string theory. The 3-brane is left invariant but the
fundamental string and D-string form a doublet along with the dilaton and
axion. Future work will concentrate on these M-theoretic applications
which require the full supersymmetric actions and also the
possibility of including other M-theory modes such as those from the
membrane. 

The structure of the paper is as follows. We will begin by reviewing the
action for Born-Infeld coupled to an axion and dilaton and the resulting
duality symmetry. We then describe how this action may be obtained
from non linear self dual 2-form electrodynamics in 6 dimensions. We then
demonstrate how by dropping Lorentz invariance we may also construct an
action for the Born-Infeld theory that duality symmetry manifest,
providing a Schwarz Sen type action for Born-Infeld. Finally, we discuss
how this fits into the brane picture.

\bigskip
\bigskip
\bigskip

{\bigtitle{Born-Infeld action and duality}}

\bigskip

We begin with the Born-Infeld action with the coupling to a
background dilaton, $\phi$
and axion, $C$.

$$S={\int}_{M^4}d^4x \sqrt{ -det(\eta_{\mu \nu} + e^{-\phi/2} F_{\mu \nu}
) }
+ {1 \over 8}i\epsilon^{\mu \nu \rho \sigma} C F_{\mu \nu} F_{\rho
\sigma} \eqno(1)$$

We will introduce a complex field $\lambda = C + i e^{-\phi}$ that will
prove useful later.  $\eta$ is the metric and $F$ the field strength of an
abelian vector potential A, defined as usual by $F=dA$. We will work in
flat space time with $\eta=diag(1,-1,-1,-1)$. We performing the duality
transformation on the above action by the usual process [5]. That is
treat $F$ as a generic two from and then add a term to the lagrangian that
imposes F be closed via a lagrange multiplier. We can then integrate out F
in favor of the lagrange multiplier in the path integral, or classically,
solve the equation of motion for F in terms of the lagrange multiplier and
then substituting into the lagrangian. The duality
transformed action is,
in terms of dual gauge field strength denoted with a tilde:

$$S_D = \int_{M^4} d^4x \sqrt{ -det(\eta_{\mu \nu} + {1 \over
{\sqrt{e^{-\phi}
+ e^{\phi} C^2}}}{\tilde{F_{\mu \nu}}} )  } + {1 \over 8}i\epsilon^{\mu
\nu
\rho \sigma} {- C e^{\phi} \over e^{-\phi} + e^{\phi} C^2} {\tilde{F_{\mu
\nu}}} {\tilde{F_{\rho \sigma}}} \eqno(2)$$ If we define the dual dilaton
axion as follows: $$e^{-\tilde{\phi}}= {1
\over e^{-\phi} +
e^{\phi} C^2}$$ and 
$$\tilde{C} = {- C e^{\phi} \over e^{-\phi} + e^{\phi} C^2}$$
Hence, the dual of $\lambda$ becomes:
$${\tilde{\lambda}}=
{-1 \over \lambda} \eqno(3)$$

The
action (2) is then identical to (1) for the dual fields. If we think 
of the background dilaton axion as playing the role of an effective gauge
coupling constant we have the following picture. The form of the action is
left invariant but the coupling is inverted. This is like the usual
S-duality.  There is also the usual allowed integer shifts in
C that leave the path integral invariant (strictly speaking we should
normalize our action by factor of $4 \pi$). Together the two
transformations generate the full SL(2,Z). 

As has been shown in a variety of contexts [8], duality
related theories can
be
obtained by compactification of a single parent theory in higher
dimensions. In
the case of
Maxwell theory in 4-dimensions, S-duality was derived by compactifying
a self dual 2-form theory Maxwell theory in six
dimensions. (By the Maxwell theory, we mean a theory with an action that 
has the usual field strength squared as opposed to the the Born-Infeld
type theories which have a more complicated non-linear action.) To
generalize this idea to the case of Born-Infeld
however,
we require that the six dimensional, parent theory
also be non-linear and self dual. The Lorentz invariant form for such a
theory was not known until very recently. Schwarz and others [7] in
the
context of
searching for the M-theory five brane action produced the action from
which we will take our starting point. 

$$S=-\int_{M^6} d^6x \sqrt{-det(G_{\mu \nu} + i { \tilde{H}_{\mu \nu}
\over
\sqrt{ (\partial a)^2} })} - { \tilde{H}^{\mu \nu} H_{\mu \nu \rho}
\partial ^{\rho} a \over 4 (\partial a)^2 } \eqno(4) $$ $G_{\mu \nu}$ is
the metric in six dimensions. We will introduce some form notation that
will be useful later. The space of p-forms that have values on a d
dimensional manifold, $M^d$ is called $\Lambda^p(M^d)$. So, for the fields
in the above action:  ${\tilde{H}} \in \Lambda^2(M^6)$, $a \in
\Lambda^0(M^6)$. We define ${\tilde{H}}$ by the following: $$\tilde{H} =
{}^*(H \wedge da) \eqno(5)$$ Where $^*$ is the Hodge dual acting in 6
dimensions.
$H \in \Lambda^3(M^6)$ is the field strength of the abelian potential $B
\in \Lambda^2(M^6)$ defined by the usual relation $H=dB$. 

The field $a$ is completely auxiliary. However, it is required to preserve
Lorentz invariance in the action. We will not discuss all the properties
of this action here but refer to the literature [7]. However, there are
two symmetries that will
prove relevant. One is the usual gauge symmetry for an abelian potential,
$\delta B = d \chi$. The other is the non trivial gauge
symmetry introduced by the new auxiliary field : 

$$ \delta B = \psi \wedge da \eqno(6)$$ where $\psi \in \Lambda^1(M^6)$
is
the gauge parameter.
\bigskip
\bigskip
{\bigtitle{Dimensional Reduction}}
\bigskip
\bigskip
Now, we will double dimensionally reduce this action on a torus, keeping
only
the zero modes. So we carry
out the following: $M^6 \rightarrow M^4 \times T^2$. We make the following
ans{\"a}tze, $$G = \eta \oplus \pi \eqno(7)$$ where $\pi$ is the metric
on
the
torus and $\eta$ the metric in four dimensions. This is in fact a
truncation (consistent) where we do not consider the possible
Kaluza-Klein fields corresponding to the compact dimensions, of
which there should be two. Our ansatz for
the gauge field $B$ is again truncated. We have only included a part that
couples to the conformal part of the torus. Hence, $$B =A^I \gamma_I \quad
\Rightarrow \quad H=F^I \gamma_I \eqno(8)$$ where $A^I \in
\Lambda^1(M^4)$, $F^I=d A^I$ and $\gamma_I$ are the canonical one forms
associated with the non trivial homology one cycles one the torus. Hence,
they form a basis for $H^1(T^2,Z)$. There are two such one cycles, hence
$I=1,2$. Note, $d \gamma_I =0$ and $d^* \gamma_I = 0$.

Now we have two natural possibilities for the auxiliary field $a$. It can
be chosen such that  $da \in \Lambda^1(T^2)$ or $da \in \Lambda^1(M^4)$.
We will look at the consequences of both choices. Though of course, both
possibilities must be physically equivalent. In the first instance, we
find the following for $\tilde{H}$: $$ \tilde{H} = {}^*F^I { } 
{^*}(
\gamma_I \wedge da)  \eqno(9)$$ where the Hodge star in front of $F$ acts
in $M^4$ and the
Hodge
star in front of the parentheses acts in $T^2$. This gives $\tilde{H} \in
\Lambda^2(M^4)$. We can now factorize the determinant,
using $det(A \oplus B) = det(A) det(B)$. So that $$S=\int_{M^4}\int_{T^2} 
-\sqrt{\pi}
\sqrt{-det(\eta_{\alpha \beta} + i { {^*}F{^I}_{\alpha \beta} {^*}(
\gamma_I \wedge da)\over
\sqrt{ (\partial a)^2} })} - { {^*}F^{I \mu \nu} F{^J}_{\mu \nu}
{^*}(\gamma_I
\wedge da){}^* (\gamma_J \wedge ^*da) \over 4 (\partial a)^2 }
\eqno(10)
$$ 

To investigate this action we will now make a gauge choice for $da$. A
natural choice is to take $da \in H^1(T^2,Z)$. So suppose we choose $da$
to
be $\gamma_L$. The local symmetry (4) then allows us to
gauge away one of the fields, $A^L$. It only remains to evaluate the 
terms in the
action such as
$\gamma_L \wedge
\gamma_I$ and $\gamma_L \wedge ^* \gamma_I$. We can evaluate these using
an explicit basis for $H^1(T^2,Z)$. These terms are proportional to the
volume
form $\Omega$ as follows:

$$\gamma_I \wedge ^*\gamma_J= {M_{IJ} \Omega \over {\cal{V}}},
\qquad \gamma_I \wedge \gamma_J={L_{IJ}  \Omega  \over {\cal{V}}}
\eqno(11a)$$ where  ${\cal{V}}=\int_{T^2} \Omega$ an
d $M_{IJ}$ and
$L_{IJ}$ are the period and
intersection matrices
defined as follows:

 $$M = \int_{T^2} \left(\matrix{ \gamma_1 \wedge ^* \gamma_1 &
\gamma_1 \wedge ^* \gamma_2 \cr  
                                    \gamma_2 \wedge ^* \gamma_1 & \gamma_2 
\wedge ^* \gamma_2 \cr}
\right)
        = {1 \over {\tau_2}}\left(\matrix{ 1 & \tau_1 \cr
                     \tau_1 & |\tau|^2 \cr}\right) ,
\quad L_{IJ}=\int_{T^2}
\gamma_I \wedge
\gamma_J = \left(\matrix{ 0 & 1 \cr -1 & 0 \cr}\right) \eqno(11b)$$ 
Hence, substituting in (11) into (10) and integrating over the
torus we find the action:

 $$S=\int_{M^4} -\sqrt{-det({\sqrt{{\cal{V}}}}\eta_{\alpha \beta} + i {
^*F_{\alpha \beta}
\omega)}} - { ^*F^{\mu \nu} F_{\mu \nu} \rho } \eqno(12)  $$ Where
$\omega$ and $\rho$ depend on the specific choice of $da$. The two
independent choices for $da$ give the following:
$$da=\gamma_1  \quad \Rightarrow \quad \omega= \sqrt{\tau_2} ,
\quad \rho=\tau_1$$
$$da=\gamma_2 \quad \Rightarrow \quad  \omega= \sqrt{\tau_2 \over
|\tau|^2} ,
\quad \rho={-\tau_1 
\over |\tau|^2}     \eqno(13)$$
Redefining, $F^{\prime} = i^*F$ and rescaling the metric as follows
${{\eta{\prime}_{\alpha \beta}}} = {\sqrt{{\cal{V}}}}\eta_{\alpha
\beta}$ allows us to identify the action (12) with the Born-Infeld action
given in (1). With this identification we then compare the action (12)
for different choices of $da$ with the actions (1) and (2).
For choice $da=\gamma_1$ we identify (12) with (1) and for $da=\gamma_2$
we identify (12) with the dual theory (2). These identifications imply
simply that we
must identify the dilaton-axion
with the complex structure. That is, $$\lambda=\tau \eqno(14)$$ The
duality transformation that
inverts $\lambda$ is then given by
making a different choice for $da$. Hence, we see how duality becomes a
gauge symmetry of this theory. 

The other possibility mentioned above is that $da \in \Lambda^1(M^4)$.
Note, that once such a
choice is made, manifest Lorentz invariance is broken as $da$ picks
out a direction in space time. We will go immediately to the obvious
choice $da= dt$. Other choices for $da$, will not be related by duality as
in the previous case but by Lorentz transformations. We use the same
ansatze as before for the metric and the two form gauge field $B$, however
now we find that the matrix, $G + i {\tilde{H}}$ does not decompose into
block diagonal form and so the determinant will not immediately
factorise. 
Hence, we explicitly expand out the determinant using the following
identity (where $H_{\mu \nu}$ is an antisymmetric tensor in 6 dimensions
of rank 4):  $$det(G_{\mu \nu} + i H_{\mu \nu})=detG (1 + {1 \over 2}trH^2
+ {1\over 8} (trH^2)^2 - {1 \over 4} tr H^4) \eqno(15)$$ We now define
the
magnetic and electric field strengths in the usual way: $B_i = {1 \over 2}
\epsilon_i{}^{jk} F_{jk}$ where $i,j=1,2,3$ and $E_i= F_{i0}$. We now
substitute in the
E and B fields into the action expanded out using the above identity. We
also have used the period and intersection matrices of the torus as
before and integrated over the torus. We also make the same scaling of the
metric so to absorb the area dependence into the metric.
$$S=\int_{M^4}d^4x \sqrt{-{\eta}\prime} \sqrt{-[1+B_i^I B^{Ji} M_{IJ} +
{1 \over
2}
B_i^I
B^{Ji} B^L _j B^{Kj} M_{IJ} M_{LK} - { 1 \over 2} B_i^I    
B^{Li} B^J _j B^{Kj} M_{IJ} M_{LK}]} $$ $$+ E^1_i B^{2i} - E^2_i B^{1i}
\eqno(16)$$ 

So note, this action with two magnetic fields is symmetric in $B^1$ and
$B^2$. (Obviously, a choice of space-like $da$ would give a pair of
electric fields.) These fields are related to each other by
duality, as we will show when we demonstrate the equivalence of the
above action to Born-Infeld. It is this action that we claim is the
Born-Infeld equivalent
of
the Schwarz-Sen action for Maxwell theory. (Recently, several people,
using very different approaches to those described here, have produced
manifest
duality actions for Born-Infeld theory [9]).

We will now go to the case where $\tau=i$ as this will ease our
calculation greatly. We will reinstate the dilaton coupling later. We will
now
follow the method of [1] Schwarz and Sen to show that this action
gives the Born-Infeld in 4-dimensions. First use gauge invariance to set
$A_0=0$. Then, as discussed in [1] one of 
the $A^L_i$ field becomes auxiliary and may be eliminated in favor of
the other. Let us work in the
concrete case where we will eliminate $A^2$ from the action (9). We find
the equation of motion for $A^2$ by varying the action (9) (with $M$
equal to the identity:

$${\vec{\nabla}} \wedge ( {\vec{M}}({B}{^1},{B}{^2})  -
\vec{E}{^1} ) =
0$$
Where$$
{\vec{M}}({B}{^1},{B}{^2})=
{\vec{B}{^2} - (\vec{B}{^2} \cdot \vec{B}{^1}) \vec{B}{^1} + (\vec{B}{^1}
\cdot \vec{B}{^1}) \vec{B}{^2} \over {\sqrt{ 1+ (\vec{B}{^1})^2
(\vec{B}{^2})^2-
(\vec{B}{^1} \cdot
\vec{B}{^2})^2  +
(\vec{B}{^1})^2 + (\vec{B}{^2})^2 }}}$$ with ${\vec{B}}$
being a vector in 3 dimensions. We can solve
this
by writing $${{\vec{M}}(B^1,B^2)} - {\vec{E}}^1 ={\vec{\nabla}} \psi$$
We still have some gauge symmetry left $\delta {\vec{A}}^1 =
{\vec{\nabla}} \chi$ to
eliminate
$\nabla
\psi$. Leaving the equation:$$ {{\vec{M}}(B^1,B^2)} - {\vec{E}}^1 = 0
\eqno(17)$$
The equivalent equation in Schwarz Sen approach to Maxwell theory is
simply ${\vec{B}}^2={\vec{E}}^1$, which greatly facillitates the
calculation and explicitly shows that the pair of Electric and Magnetic
fields are
related by duality.

The next step is to solve this equation for ${\vec{B}}^2$. After some
manipulations we find $${\vec{B}}^2= { {\vec{E}}^1 +( {\vec{E}}^1 \cdot
{\vec{B}}^1)
{\vec{B}}^1 \over {\sqrt{ 1+ ({\vec{B}}^1)^2
({\vec{B}}{^2})^2-
(\vec{B}{^1} \cdot
\vec{B}{^2})^2  +
(\vec{B}{^1})^2 + (\vec{B}{^2})^2}}} \eqno(18)$$ As a simple check we can
see that this equation for $B^2$ reduces to to Maxwell case to first order
in fields. 

We now substitute this into the action (9) and find:
$$S=\int_{M^4}d^4x \sqrt{-\eta{\prime}} \sqrt{(1+ ({\vec{B}}^1)^2 -
({\vec{E}}^1)^2
- ({\vec{E}}^1 \cdot
{\vec{B}}^1)^2)}$$
This becomes after rewriting in terms of a four dimensional determinant:
$$S= \int_{M^4}d^4x
\sqrt{-det(\eta{\prime}_{\mu \nu} + F_{\mu \nu} ) }     \eqno(19)$$

This is of course the Born-Infeld with trivial background fields. If we
reinstate the dilaton coupling and repeat the above procedure we see that
we get the expected dilaton dependence. That is, we recover the action
given in equation (1) without the axion term. We have so far been unable
to repeat the process with the axion term, essentially because we have
not been able to solve the analogue of equation (17) once the axion is 
included. There is no reason to believe that it can not be done and this
detail would not add anything to the overall picture.

We generate the dual
theory by repeating the process but instead we integrate out $A^1$ instead
of $A^2$. This gives the same action but with the expected
dilaton inversion. So in this
description the duality is a symmetry of the action (16). The two duality
related theories are given by eliminating different fields from this
action. It is
a nice check that the two routes, one with $da$ in the compact space and
one with $da$ in space time give (as they obviously should) the same
results.

\bigskip\bigskip {\bigtitle{Discussion}} \bigskip

We can interpret our results in the context of M-theory as follows. We
will work with bosonic branes. To interpret the actions presented here as
branes we write the metric as a pull-back onto the brane from the
background space-time. That is: 

$$G_{\mu \nu} = \partial_\mu X^M \partial_\nu X^N g_{MN}  \eqno(20)$$
Where $X^M$ are D dimensional spacetime coordinates, with $g_{MN}$ the
D-dimensional spacetime metric. $\partial_\mu = { \partial \over {\partial
x^\mu}}$ where $x^\mu$ are coordinates in the d-brane, $\mu=0..d$. The
dimensional reduction described above becomes double dimensional
reduction in the brane picture. That is, we identify the compact brane
coordinate with the compact space-time coordinate. Now
we would like to justify our ans{\"a}tze used for the dimensional
reduction
from this point of view. There are two spacetime U(1) gauge fields from
Kaluza-Klein on the torus. The three-brane momentum in the compact
dimensions couples to
these gauge fields. By taking the zero modes on the torus we are
identifying the sector of the theory in which the 3-brane is neutral with
respect to these fields. Hence, the truncation of these
fields.

By not considering the scalar fields and two form fields in our ansatz
for $B$ we are truncating the part that is coupled to the area of the
torus. (As $H$ is self dual these fields will be duality related). We
justify this as follows.

We wish to compare our double dimensional reduction from 11 to 9
dimensions of a 5-brane with the direct dimensional reduction of a 3-brane
from 10 to 9 dimensions. Direct reduction implies that we do not wrap the
brane around the compact space-time dimension. This induces an additional
scalar field in the brane with a coupling given by
the radius squared of the compact dimension. In the spirit of [10] the
world volume dual of this field will then be identified with the two form
potential $B$ in the 5-brane picture. In doing this identification we must
identify the radius, $R$ of the 10th dimension with ${1 \over
{\sqrt{{\cal{V}}}}}$. With this identification can then see that our 
metric $\eta$ is scaled by $ 1 \over R$. This is what is to be expected
following arguments given in [11] to allow $\eta\prime$ to be
identified with the 10-dimensional metric in the Einstien frame (it is in
this metric that the IIB 3-brane is self dual). If we consider the limit
where the area
of the torus goes to zero (keeping the complex structure fixed) we must
go to the limit where $R$ goes to
infinity and lift to a non compact 10-dimensional theory. Hence, by
truncating the fields that couple to the area of the torus we are going
immediately to the description of the 3-brane in 10 dimensions. The
supersymmetric version of these actions will be interpreted as the IIB
D-3 brane from the double dimensional reduction of
the M 5-brane on a torus. The IIB S-duality then becomes manifest as
the modular group of the torus, as reported in [12]. Our
S-duality in the brane is then a result of the IIB S-duality. We will
report
in the future the more precise M-theory, brane picture that extends the
above ansatz and includes the important Ramond-Ramond fields.  

\bigskip
\bigskip
{\bigtitle{Conclusions}}
\bigskip
We confirm that the duality related Born-Infeld theories with coupling to
a background dilaton-axion can come naturally, from compactification of a
self dual non-linear two form theory in 6 dimensions. The axion-dilaton
becomes identified with the complex structure of the torus, and the
duality symmetry then becomes associated with the modular group of the
torus. This is as expected from the M-theory point of view where the one
imagines the 3 brane as coming from wrapping the 5-brane on a torus.

\bigskip
{\bigtitle{Acknowledgements}}
\bigskip
This work was supported by PPARC. We wish to thank John Schwarz for
clarifying the possible applications to M-theory and reviewing the
manuscript and also thank D. Fairlie for general support.
\bigskip\bigskip

{\bigtitle{References}}
\bigskip
\bigskip

[1] J. Schwarz and A. Sen, Nuc. Phys. {\bf{B411}}, (1994) 35.

[2] R. Floreanini and R. Jackiw, Phys. Rev. Lett. {\bf{59}} (1987) 1873;

A. Tseytlin, Phys. Lett {\bf{B242}} (1990) 163

M. Henneaux and C. Teitelboim, Phys. Lett. {\bf{B206}} (1988) 650.

[3] P. Pasti, D. Sorokin and M. Tonin, Phys. Rev. {\bf{D52}} (1995) 4277;

P. Pasti, D. Sorokin and M. Tonin, Phys. Lett. {\bf{B352}} (1995) 59.

[4] J. Polchinski, "Tasi Lectures on D-branes" hep-th/9611050.

[5] Mina Aganagic, Jaemo Park, Costin Popescu, John H. Schwarz,

"Dual D-brane actions", hep-th/9701166;

A. Tseytlin, Nuc. Phys. {\bf{B469}} (1996) 51.

[6] G. Gibbons and D. Rasheed, Nuc. Phys. {\bf{B454}} (1995); Phys. Lett.
{\bf{365}} (1996).

[7] M. Perry and J. Schwarz, Nuc. Phys. {\bf{B489}} (1997) 47;

M. Aganagic, J. Park, C. Popescu, J. Schwarz, "World Volume Action of the

M-theory five brane", hep-th/9701166;

P. Pasti, D. Sorokin and M. Tonin, Phys. Lett. {\bf{B398}} (1997), 41.

I. Bandos, K. Lechner, A. Nurmagambetov, P. Pasti, D. Sorokin and M.
Tonin,

Phys. Rev. Lett. {\bf{78}} (1997), 4332-4334;

[8] D. Berman, Phys. Lett. {\bf{B403}} (1997) 250;

N.Berkovits, Phys. Lett. {\bf{B388}} (1996) 743;

A. Giveon and M. Porrati, Phys. Lett. {\bf{B385}} (1996) 81;

I. Giannakis, V. P. Nair, "Symplectic Structures and Self-dual Fields in

(4k+2) Dimensions", hep-th/9702024;

E. Verlinde, Nuc. Phys. {\bf{B455}} (1995) 211.

[9] I. Bengtsson, "Manifest Duality in Born-Infeld Theory", 
hep-th/9612174;

M. Gaillard and B. Zumino, "Self-Duality in Nonlinear 

Electromagnetism", hep-th/97052206.

[10] P. Townsend, Phys. Lett. {\bf{B373}} (1996) 68.

[11] E. Witten, Nuc. Phys. {\bf{B443}} (1995) 85.

[12] J. Schwarz, Phys. Lett. {\bf{B360}} (1995), erratum ibid {\bf{B364}}
(1995).

 \end